\documentclass[twocolumn,10pt]{article}

\usepackage{latexsym}
\usepackage{theorem}
\usepackage{amssymb}
\usepackage{amsmath}
\usepackage{graphics}

\newcommand{\comment}[1]{}
\newcommand{\bu}{$\bullet$}
\newcommand{\hs}[1][3ex]{\hspace*{#1}}
\newcommand{\vs}[1][1mm]{\vspace*{#1}}
\newcommand{\moins}{\setminus}
\newcommand{\vide}{\emptyset}

\renewcommand{\l}[2]{[#1\!:\!#2]}
\newcommand{\p}[2]{(#1\!:\!#2)}
\newcommand{\lx}{\l{x}}
\newcommand{\px}{\p{x}}

\newcommand{\pz}{\p{z}}

\newcommand{\pX}{\p{X}}

\newcommand{\dom}{\mi{dom}}

\newcommand{\FV}{\mi{FV}}
\newcommand{\pos}{\mi{Pos}}

\renewcommand{\a}{\rightarrow}
\newcommand{\A}{\Rightarrow}
\renewcommand{\aa}{\leftrightarrow}

\newcommand{\ad}{\downarrow}

\renewcommand{\to}{\mapsto}

\newcommand{\ab}{\a_\b}

\newcommand{\ar}{\a_\cR}

\newcommand{\ps}[1]{{\langle #1\rangle}}

\newcommand{\I}[1]{[\![#1]\!]}
\newcommand{\n}[1]{|#1|}

\newcommand{\ex}{\exists}
\newcommand{\all}{\forall}
\newcommand{\ou}{\vee}
\newcommand{\et}{\wedge}
\newcommand{\non}{\neg}
\newcommand{\st}{\star}
\newcommand{\B}{\Box} % latexsym
\renewcommand{\th}{\vdash}

\newcommand{\IN}{\!\in\!}

\newcommand{\sle}{\subseteq}

\newcommand{\tgt}{\rhd}

\renewcommand{\o}[1]{{\overline{#1}}}

\renewcommand{\b}{\beta}
\newcommand{\g}{\gamma}
\newcommand{\G}{\Gamma}
\renewcommand{\d}{\delta}
\newcommand{\D}{\Delta}

\newcommand{\vep}{\varepsilon}

\renewcommand{\t}{\theta}
\newcommand{\T}{\Theta}
\renewcommand{\i}{\iota}
\renewcommand{\k}{\kappa}
\newcommand{\la}{\lambda}
\renewcommand{\L}{\Lambda}
\renewcommand{\r}{\rho}
\newcommand{\s}{\sigma}

\newcommand{\mi}{\mathit}
\newcommand{\mc}{\mathcal}

\newcommand{\mr}{\mathrm}

\newcommand{\cC}{\mc{C}}
\newcommand{\cD}{\mc{D}}

\newcommand{\cF}{\mc{F}}
\newcommand{\cG}{\mc{G}}

\newcommand{\cR}{\mc{R}}
\newcommand{\cS}{\mc{S}}

\newcommand{\cX}{\mc{X}}

\newenvironment{rew}%
  {$\begin{array}{r@{~\a~}l}}%
  {\end{array}$}
\newenvironment{rewc}%
  {\begin{center}\begin{rew}}%
  {\end{rew}\end{center}}

\newcounter{counter}

{\theorembodyfont{\rmfamily} % theorem
  \newtheorem{dfn}[counter]{Definition}
  
  \newtheorem{thm}[counter]{Theorem}

}

\newenvironment{lstgeneric}[2]
  {\begin{list}{#1}{\topsep=0ex\itemsep=0ex\parsep=0ex%
    \itemindent=-3ex\labelsep=1ex\labelwidth=0ex #2}}
  {\end{list}}
\newenvironment{lst}[1]
  {\begin{lstgeneric}{#1}{\itemindent=-1ex}}
  {\end{lstgeneric}}
\newenvironment{enumi}[1]
  {\begin{lstgeneric}{}{\usecounter{enumi}\leftmargin=7mm%
    }}
  {\end{lstgeneric}}
\newenvironment{bfenumi}[1]
  {\begin{lstgeneric}{}{\usecounter{enumi}\leftmargin=7mm%
    }}
  {\end{lstgeneric}}

\begin{document}

% input files
%%%%%%%%%%%%%%%%%%%%%%%%%%%%%%%%%%%%%%%%%%%%%%%%%%%%%%%%%%%%%%%%%%%%%%%%%%%%%%

% title

%%%%%%%%%%%%%%%%%%%%%%%%%%%%%%%%%%%%%%%%%%%%%%%%%%%%%%%%%%%%%%%%%%%%%%%%%%%%%%

\title{\bf Definitions by Rewriting\\
in the Calculus of Constructions}

\author{Fr\'ed\'eric Blanqui\vspace*{1mm}\\
\small LRI, b\^at. 490, Universit\'e Paris-Sud, 91405 Orsay Cedex, France\\
\small tel: +33 (0) 1 69 15 42 35 ~~~fax: +33 (0) 1 69 15 65 86\\
%\small {\tt http://www.lri.fr/\~{}blanqui/}\\
\small {\tt blanqui@lri.fr}
}

\date{}

\maketitle

\noindent {\bf Abstract~:}
{\it The main novelty of this paper is to consider an extension of the
  Calculus of Constructions where predicates can be defined with a
  general form of rewrite rules.
  
  We prove the strong normalization of the reduction relation
  generated by the $\b$-rule and the user-defined rules under some
  general syntactic conditions including confluence.
  
  As examples, we show that two important systems satisfy these
  conditions~: a sub-system of the Calculus of Inductive Constructions
  which is the basis of the proof assistant Coq, and the Natural
  Deduction Modulo a large class of equational theories.}

%%% Local Variables: 
%%% mode: latex
%%% TeX-master: "main"
%%% End: 

%%%%%%%%%%%%%%%%%%%%%%%%%%%%%%%%%%%%%%%%%%%%%%%%%%%%%%%%%%%%%%%%%%%%%%%%%%%%%%
%%%%%%%%%%%%%%%%%%%%%%%%%%%%%%%%%%%%%%%%%%%%%%%%%%%%%%%%%%%%%%%%%%%%%%%%%%%%%%
%%%%%%%%%%%%%%%%%%%%%%%%%%%%%%%%%%%%%%%%%%%%%%%%%%%%%%%%%%%%%%%%%%%%%%%%%%%%%%

% introduction

%%%%%%%%%%%%%%%%%%%%%%%%%%%%%%%%%%%%%%%%%%%%%%%%%%%%%%%%%%%%%%%%%%%%%%%%%%%%%%
%%%%%%%%%%%%%%%%%%%%%%%%%%%%%%%%%%%%%%%%%%%%%%%%%%%%%%%%%%%%%%%%%%%%%%%%%%%%%%
%%%%%%%%%%%%%%%%%%%%%%%%%%%%%%%%%%%%%%%%%%%%%%%%%%%%%%%%%%%%%%%%%%%%%%%%%%%%%%

\section{Introduction}

This work aims at defining an expressive language allowing to specify
and prove mathematical properties in which functions and predicates
can be defined by rewrite rules, hence enabling the automatic proof of
equational problems.

\vs[2mm]
\noindent
{\bf The Calculus of Constructions.} The quest for such a language
started with Girard's system F \cite{girard88book} on one hand and De
Bruijn's Automath project \cite{debruijn94book} on the other hand.
Later, Coquand and Huet combined both calculi into the Calculus of
Constructions (CC) \cite{coquand88ic}. As in system F, in CC, data
structures are defined by using an impredicative encoding which is
difficult to use in practice. Following Martin-L\"of's theory of types
\cite{martinlof84book}, Coquand and Paulin-Mohring defined an
extension of CC with inductive types and their associated induction
principles as first-class objects~: the Calculus of Inductive
Constructions (CIC) \cite{paulin93tlca} which is the basis of the
proof-assistant Coq \cite{coq}.

\vs[2mm]
\noindent
{\bf Reasoning Modulo.} Defining functions or predicates by recursion
is not always convenient. Moreover, with such definitions, equational
reasoning is uneasy and leads to very large proof terms. Yet, for
decidable theories, equational proofs need not to be kept in proof
terms. This idea that proving is not only reasoning (undecidable) but
also computing (decidable) has been recently formalized in a general
way by Dowek, Hardin and Kirchner with the Natural Deduction Modulo
(NDM) for first-order logic \cite{dowek98trtpm}.

\vs[2mm]
\noindent
{\bf Object-level rewriting.} In CC, the first extension by a general
notion of rewriting is the $\la R$-cube of Barbanera, Fern\'andez and
Geuvers \cite{barbanera97jfp}. Their work extends the works of
Breazu-Tannen and Gallier \cite{breazu91tcs} and Jouannaud and Okada
\cite{jouannaud97tcs} on the combination of typed $\la$-calculi with
rewriting. The notion of rewriting considered in
\cite{jouannaud97tcs,barbanera97jfp} is not restricted to first-order
rewriting, but also includes higher-order rewriting following
Jouannaud and Okada's General Schema \cite{jouannaud97tcs}, a
generalization of the primitive recursive definition schema. This
schema has been reformulated and enhanced so as to deal with
definitions on strictly-positive inductive types \cite{blanqui99rta}
and with higher-order pattern-matching \cite{blanqui00rta}.

\vs[2mm]
\noindent
{\bf Predicate-level rewriting.} The notion of rewriting considered in
\cite{barbanera97jfp} is restricted to the object-level while, in CIC
or NDM, it is possible to define predicates by recursion or by
rewriting respectively. Recursion at the predicate-level is called
``strong elimination'' in \cite{paulin93tlca} and has been shown
consistent by Werner \cite{werner94thesis}.

\vs[2mm]
\noindent
{\bf Our contributions.} The main contribution of our work is a strong
normalization result for the Calculus of Constructions extended with,
at the predicate-level, user-defined rewrite rules satisfying some
general admissibility conditions. As examples, we show that these
conditions are satisfied by a sub-system of CIC with strong
elimination \cite{paulin93tlca} and the Natural Deduction Modulo
\cite{dowek98types} a large class of equational theories.

So, our work can be used as a foundation for an extension of a proof
assistant like Coq \cite{coq} where users could define functions and
predicates by rewrite rules. Checking the admissibility conditions or
the convertibility of two expressions may require the use of external
specialized tools like CiME \cite{cime} or ELAN \cite{elan}.

\vs[2mm]
\noindent
{\bf Outline of the paper.} In Section~\ref{sec-cac}, we introduce the
Calculus of Algebraic Constructions and our notations. In
Section~\ref{sec-admi}, we present our general syntactic conditions.
In Section~\ref{sec-examples}, we apply our result to CIC and NDM.  In
Section~\ref{sec-conclu}, we summarize the main contributions of our
work and, in Section~\ref{sec-future}, we give future directions of
work. Detailed proofs can be found in \cite{blanqui01thesis}.

%%% Local Variables: 
%%% mode: latex
%%% TeX-master: "main"
%%% End: 

%%%%%%%%%%%%%%%%%%%%%%%%%%%%%%%%%%%%%%%%%%%%%%%%%%%%%%%%%%%%%%%%%%%%%%%%%%%%%%
%%%%%%%%%%%%%%%%%%%%%%%%%%%%%%%%%%%%%%%%%%%%%%%%%%%%%%%%%%%%%%%%%%%%%%%%%%%%%%
%%%%%%%%%%%%%%%%%%%%%%%%%%%%%%%%%%%%%%%%%%%%%%%%%%%%%%%%%%%%%%%%%%%%%%%%%%%%%%

% the calculus of algebraic constructions

%%%%%%%%%%%%%%%%%%%%%%%%%%%%%%%%%%%%%%%%%%%%%%%%%%%%%%%%%%%%%%%%%%%%%%%%%%%%%%
%%%%%%%%%%%%%%%%%%%%%%%%%%%%%%%%%%%%%%%%%%%%%%%%%%%%%%%%%%%%%%%%%%%%%%%%%%%%%%
%%%%%%%%%%%%%%%%%%%%%%%%%%%%%%%%%%%%%%%%%%%%%%%%%%%%%%%%%%%%%%%%%%%%%%%%%%%%%%

\section{The Calculus of Algebraic Constructions (CAC)}
\label{sec-cac}

%%%%%%%%%%%%%%%%%%%%%%%%%%%%%%%%%%%%%%%%%%%%%%%%%%%%%%%%%%%%%%%%%%%%%%%%%%%%%%
% macros
%%%%%%%%%%%%%%%%%%%%%%%%%%%%%%%%%%%%%%%%%%%%%%%%%%%%%%%%%%%%%%%%%%%%%%%%%%%%%%

\newcommand{\FVB}{\FV^\B}

\newcommand{\vx}{\vec{x}}
\newcommand{\vt}{\vec{t}}
\newcommand{\vT}{\vec{T}}
\newcommand{\vy}{\vec{y}}
\newcommand{\vU}{\vec{U}}
\newcommand{\vl}{\vec{l}}
\newcommand{\vu}{\vec{u}}
\newcommand{\vz}{\vec{z}}
\newcommand{\vV}{\vec{V}}
\newcommand{\vv}{\vec{v}}
\newcommand{\va}{\vec{a}}
\newcommand{\vw}{\vec{w}}
\newcommand{\vS}{\vec{S}}

\newcommand{\pvxT}{(\vx:\vT)}
\newcommand{\pvxt}{(\vx:\vt)}
\newcommand{\pvxU}{(\vx:\vU)}
\newcommand{\pvyU}{(\vy:\vU)}
\newcommand{\pvzV}{(\vz:\vV)}

\newcommand{\xt}{\{x\to t\}}
\newcommand{\xu}{\{x\to u\}}
\newcommand{\xup}{\{x\to u'\}}
\newcommand{\xv}{\{x\to v\}}
\newcommand{\yv}{\{y\to v\}}
\newcommand{\XU}{\{X\to U\}}
\newcommand{\XS}{\{X\to S\}}
\newcommand{\xS}{\{x\to S\}}

\newcommand{\vyu}{\{\vy\to\vu\}}
\newcommand{\vxt}{\{\vx\to\vt\}}
\newcommand{\vxl}{\{\vx\to\vl\}}
\newcommand{\vxS}{\{\vx\to\vS\}}
\newcommand{\vxu}{\{\vx\to\vu\}}

\newcommand{\tf}{\tau_f}
\newcommand{\tg}{\tau_g}
\newcommand{\tF}{\tau_F}
\newcommand{\tG}{\tau_G}
\newcommand{\tc}{\tau_c}
\newcommand{\td}{\tau_d}
\newcommand{\tC}{\tau_C}

\renewcommand{\c}{~\cC~}

\newcommand{\E}{\mathbb{E}}
\renewcommand{\T}{\mathbb{T}}
\newcommand{\C}{\mathbb{C}}
\newcommand{\K}{\mathbb{K}}
\renewcommand{\P}{\mathbb{P}}
\renewcommand{\O}{\mathbb{O}}

\newcommand{\GxT}{\G, x\!:\!T}
\newcommand{\GxU}{\G, x\!:\!U}
\newcommand{\GxV}{\G, x\!:\!V}
\newcommand{\GXK}{\G, X\!:\!K}

\newcommand{\GpxU}{\G', x\!:\!U}
\newcommand{\GpxT}{\G', x\!:\!T}

\newcommand{\cv}[1][]{~\cC_{#1}^*~}
\newcommand{\cvG}{\cv[\G]}
\newcommand{\CV}[1][]{~\C_{#1}^*~}
\newcommand{\CVG}{\CV[\G]}

\newcommand{\oT}{\o{\T}}

\newcommand{\ti}{\tau^i}

\newcommand{\at}{\alpha}

%%%%%%%%%%%%%%%%%%%%%%%%%%%%%%%%%%%%%%%%%%%%%%%%%%%%%%%%%%%%%%%%%%%%%%%%%%%%%%
%%%%%%%%%%%%%%%%%%%%%%%%%%%%%%%%%%%%%%%%%%%%%%%%%%%%%%%%%%%%%%%%%%%%%%%%%%%%%%
%%%%%%%%%%%%%%%%%%%%%%%%%%%%%%%%%%%%%%%%%%%%%%%%%%%%%%%%%%%%%%%%%%%%%%%%%%%%%%

% syntax

%%%%%%%%%%%%%%%%%%%%%%%%%%%%%%%%%%%%%%%%%%%%%%%%%%%%%%%%%%%%%%%%%%%%%%%%%%%%%%
%%%%%%%%%%%%%%%%%%%%%%%%%%%%%%%%%%%%%%%%%%%%%%%%%%%%%%%%%%%%%%%%%%%%%%%%%%%%%%
%%%%%%%%%%%%%%%%%%%%%%%%%%%%%%%%%%%%%%%%%%%%%%%%%%%%%%%%%%%%%%%%%%%%%%%%%%%%%%

\subsection{Syntax and notations}

We assume the reader familiar with the basics of rewriting
\cite{dershowitz90book} and typed $\la$-calculus
\cite{barendregt92book}.

%%%%%%%%%%%%%%%%%%%%%%%%%%%%%%%%%%%%%%%%%%%%%%%%%%%%%%%%%%%%%%%%%%%%%%%%%%%%%%
% terms
%%%%%%%%%%%%%%%%%%%%%%%%%%%%%%%%%%%%%%%%%%%%%%%%%%%%%%%%%%%%%%%%%%%%%%%%%%%%%%

\vs[2mm]
\noindent
{\bf Sorts and symbols.} Throughout the paper, we let $\cS =
\{\st,\B\}$ be the set of {\em sorts\,} where $\st$ denotes the
impredicative universe of propositions and $\B$ a predicative universe
containing $\st$. We also assume given a family $\cF =
(\cF^s_n)^{s\in\cS}_{n\ge 0}$ of sets of {\em symbols\,} and a family
$\cX = (\cX^s)^{s\in\cS}$ of infinite sets of {\em variables\,}. A
symbol $f \in \cF^s_n$ is said to be of {\em arity\,} $\at_f = n$ and
sort $s$. $\cF^s$, $\cF_n$, $\cF$ and $\cX$ respectively denote the
set of symbols of sort $s$, the set of symbols of arity $n$, the set
of all symbols and the set of all variables.

\vs[2mm]
\noindent
{\bf Terms.} The {\em terms\,} of the corresponding CAC are given by
the following syntax~:

\begin{center}
$t ::= s ~|~ x ~|~ f(\vt) ~|~ (x:t)t ~|~ [x:t]t ~|~ tt$
\end{center}

\noindent
where $s \in \cS$, $x \in \cX$ and $f$ is applied to a vector $\vt$ of
$n$ terms if $f\in\cF_n$. $\lx{U}t$ is the abstraction and $\px{U}V$
is the product. A term is {\em algebraic\,} if it is a variable or of
the form $f(\vt)$ with each $t_i$ algebraic.

\vs[2mm]
\noindent
{\bf Notations.} As usual, we consider terms up to
$\alpha$-conversion. We denote by $\FV(t)$ the set of free variables
of $t$, by $\FV^s(t)$ the set $\FV(t) \cap \cX^s$, by $t\xu$ the term
obtained by substituting in $t$ every free occurrence of $x$ by $u$,
by $\dom(\t)$ the domain of the substitution $\t$, by $\dom^s(\t)$ the
set $\dom(\t) \cap \cX^s$, by $\pos(t)$ the set of positions in $t$
(words on the alphabet of positive integers), by $t|_p$ the subterm of
$t$ at position $p$, by $t[u]_p$ the term obtained by replacing $t|_p$
by $u$ in $t$, and by $\pos(f,t)$ and $\pos(x,t)$ the sets of
positions in $t$ where $f$ occurs and $x$ freely occurs respectively.
As usual, we write $T\a U$ for a product $\px{T}U$ where $x\notin
\FV(U)$.

%%%%%%%%%%%%%%%%%%%%%%%%%%%%%%%%%%%%%%%%%%%%%%%%%%%%%%%%%%%%%%%%%%%%%%%%%%%%%%
% rewrite rules
%%%%%%%%%%%%%%%%%%%%%%%%%%%%%%%%%%%%%%%%%%%%%%%%%%%%%%%%%%%%%%%%%%%%%%%%%%%%%%

\vs[2mm]
\noindent
{\bf Rewriting.} We assume given a set $\cR$ of {\em rewrite rules\,}
defining the symbols in $\cF$. The rules we consider are pairs $l \a
r$ made of two terms $l$ and $r$ such that $l$ is an algebraic term of
the form $f(\vl)$ and $\FV(r) \sle \FV(l)$. They induce a rewrite
relation $\ar$ on terms defined by $t \ar t'$ iff there are $p \in
\pos(t)$, $l\a r \in \cR$ and a substitution $\s$ such that $t|_p =
l\s$ and $t' = t[r\s]_p$ (matching is first-order). So, $\cR$ can be
seen as a particular case of Combinatory Reduction System (CRS)
\cite{klop93tcs} (translate $\lx{T}u$ into $\L(T,[x]u)$ and $\px{T}U$
into $\Pi(T,[x]U)$) for which higher-order pattern-matching is not
necessary.

\vs[2mm]
\noindent
{\bf Reduction.} The {\em reduction relation\,} of the calculus is $\a
\,=\, \ar \cup \ab$ where $\ab$ is defined as usual by $\lx{T}u~t$
$\ab$ $u\xt$. We denote by $\a^*$ its reflexive and transitive
closure, by $\aa^*$ its symmetric, reflexive and transitive closure,
and by $t \ad^* u$ the fact that $t$ and $u$ have a common reduct.

%%%%%%%%%%%%%%%%%%%%%%%%%%%%%%%%%%%%%%%%%%%%%%%%%%%%%%%%%%%%%%%%%%%%%%%%%%%%%%
%%%%%%%%%%%%%%%%%%%%%%%%%%%%%%%%%%%%%%%%%%%%%%%%%%%%%%%%%%%%%%%%%%%%%%%%%%%%%%
%%%%%%%%%%%%%%%%%%%%%%%%%%%%%%%%%%%%%%%%%%%%%%%%%%%%%%%%%%%%%%%%%%%%%%%%%%%%%%

% typing

%%%%%%%%%%%%%%%%%%%%%%%%%%%%%%%%%%%%%%%%%%%%%%%%%%%%%%%%%%%%%%%%%%%%%%%%%%%%%%
%%%%%%%%%%%%%%%%%%%%%%%%%%%%%%%%%%%%%%%%%%%%%%%%%%%%%%%%%%%%%%%%%%%%%%%%%%%%%%
%%%%%%%%%%%%%%%%%%%%%%%%%%%%%%%%%%%%%%%%%%%%%%%%%%%%%%%%%%%%%%%%%%%%%%%%%%%%%%

\subsection{Typing}

\noindent
{\bf Types of symbols.} We assume given a function $\tau$ which, to
each symbol $f$, associates a term $\tf$, called its {\em type\,}, of
the form $\pvxT U$ with $\n\vx = \at_f$. In contrast with our own
previous work \cite{blanqui99rta} or the work of Barbanera,
Fern\'andez and Geuvers \cite{barbanera97jfp}, symbols can have
polymorphic as well as dependent types, as it is the case in CIC.

\vs[2mm]
\noindent
{\bf Typing.} An {\em environment\,} $\G$ is an ordered list of pairs
$x_i \!:\!  T_i$ saying that $x_i$ is of type $T_i$. The {\em typing
  relation\,} of the calculus, $\th$, is defined by the rules of
Figure~\ref{fig-typing-rules} (where $s,s'\in\cS$).

\begin{figure}[ht]
\begin{center}
\caption{Typing rules\label{fig-typing-rules}}
\begin{tabular}{r@{~}c}
{\small(ax)} & $\cfrac{}{\th \st : \B}$\\

\\{\small(symb)} & $\cfrac{
\begin{array}{c}
f\in \cF^s_n,\, \tf = \pvxT U,\, \g = \vxt\\
\th \tf:s \quad \G\th v:V \quad \all i,\, \G \th t_i:T_i\g\\
\end{array}
}{\G \th f(\vt) : U\g}$\\

\\{\small(var)} & $\cfrac{\G \th T : s \quad x \in \cX^s\moins\dom(\G)}
{\GxT \th x:T}$\\

\\{\small(weak)} & $\cfrac{\G \th t : T \quad \G \th U : s \quad
x \in \cX^s\moins\dom(\G)}{\GxU \th t:T}$\\

\\{\small(prod)} & $\cfrac{\G \th T:s \quad \GxT \th U:s'}
  {\G \th \px{T}U : s'}$\\

\\{\small(abs)} & $\cfrac{\GxT \th u:U \quad \G \th \px{T}U : s}
  {\G \th \lx{T}u : \px{T}U}$\\

\\{\small(app)} & $\cfrac{\G \th t : \px{U}V \quad \G \th u : U}
  {\G \th tu : V\xu}$ \\

\\{\small(conv)} & $\cfrac{\G \th t : T \quad T \ad^* T' \quad
\G \th T' : s'}{\G \th t : T'}$\\
\end{tabular}
\end{center}
\end{figure}

An environment is {\em valid\,} if there is a term typable in it. The
condition $\G\th v:V$ in the (symb) rule insures that $\G$ is valid in
the case where $n=0$.

\vs[2mm]
\noindent
{\bf Substitutions.} Given two valid environments $\G$ and $\D$, a
substitution $\t$ is a {\em well-typed substitution\,} from $\G$ to
$\D$, written $\t : \G \a \D$, if, for all $x \in \dom(\G)$, $\D \th
x\t : x\G\t$, where $x\G$ denotes the type associated to $x$ in $\G$.
With such a substitution, if $\G \th t:T$ then $\D \th t\t:T\t$.

\vs[2mm]
\noindent
{\bf Logical consistency.} As usual, the logical consistency of such a
system is proved in three steps.

First, we must make sure that the reduction relation is correct w.r.t.
the typing relation~: if $\G \th t:T$ and $t \a t'$ then $\G \th
t':T$. This property, called {\em subject reduction\,}, is not easy to
prove for extensions of CC \cite{werner94thesis,barbanera97jfp}. In
the following subsection, we give sufficient conditions for it.

The second step is to prove that the reduction relation $\a$ is weakly
or strongly normalizing, hence that every well-typed term has a normal
form. Together with the confluence, this implies the decidability of
the typing relation which is essential in proof assistants. In this
paper, we will study the strong normalization property.

The third step is to make sure that there is no normal proof of $\bot=
\p{P}\st P$ in the empty environment. Indeed, if $\bot$ is provable
then any proposition $P$ is provable. We will not address this problem
here.

%%%%%%%%%%%%%%%%%%%%%%%%%%%%%%%%%%%%%%%%%%%%%%%%%%%%%%%%%%%%%%%%%%%%%%%%%%%%%%
%%%%%%%%%%%%%%%%%%%%%%%%%%%%%%%%%%%%%%%%%%%%%%%%%%%%%%%%%%%%%%%%%%%%%%%%%%%%%%
%%%%%%%%%%%%%%%%%%%%%%%%%%%%%%%%%%%%%%%%%%%%%%%%%%%%%%%%%%%%%%%%%%%%%%%%%%%%%%

% subject reduction

%%%%%%%%%%%%%%%%%%%%%%%%%%%%%%%%%%%%%%%%%%%%%%%%%%%%%%%%%%%%%%%%%%%%%%%%%%%%%%
%%%%%%%%%%%%%%%%%%%%%%%%%%%%%%%%%%%%%%%%%%%%%%%%%%%%%%%%%%%%%%%%%%%%%%%%%%%%%%
%%%%%%%%%%%%%%%%%%%%%%%%%%%%%%%%%%%%%%%%%%%%%%%%%%%%%%%%%%%%%%%%%%%%%%%%%%%%%%

\subsection{Subject reduction}
\label{subsec-sr}

Proving subject reduction for $\ab$ requires the following property
\cite{blanqui01thesis}~:

\begin{center}
$\px{U}V \aa^* \px{U'}V' ~\A~ U \aa^* U' ~\et~ V \aa^* V'$
\end{center}

It is easy to see that this property is satisfied when $\a$ is
confluent, an assumption which is part of our admissibility conditions
described in the next section.

\newcommand{\pl}{\p{\ell}}
\newcommand{\pA}{\p{A}}
\newcommand{\li}[1][A]{list(#1)}

For $\ar$, the idea present in all previous works is to require that,
for each rule $l\a r$, there is an environment $\G$ and a type $T$
such that $\G \th l:T$ and $\G \th r:T$. However, this approach has an
important drawback~: in presence of dependent or polymorphic types, it
leads to non-left-linear rules.

For example, consider the type $list:\st\a\st$ of polymorphic lists
built from $nil:\p{A}\st list(A)$ and $cons:\p{A}\st A\a list(A)\a
list(A)$, and the concatenation function $app:\p{A}\st list(A)\a
list(A)\a list(A)$. To fulfill the previous condition, we must define
$app$ as follows~:

\begin{rewc}
app(A,nil(A),\ell) & \ell\\
app(A,cons(A,x,\ell),\ell') & cons(A,x,app(A,\ell,\ell'))\\
\end{rewc}

This has two important consequences. The first one is that rewriting
is slowed down because of numerous equality tests. The second one is
that it may become much more difficult to prove the confluence of the
rewrite relation and of its combination with $\ab$.

We are going to see that we can take the following left-linear
definition without loosing the subject reduction property~:

\begin{rewc}
app(A,nil(A'),\ell) & \ell\\
app(A,cons(A',x,\ell),\ell') & cons(A,x,app(A,\ell,\ell'))\\
\end{rewc}

Let $l=app(A,cons(A',x,\ell),\ell')$, $r=cons(A,x,$
$app(A,\ell,\ell'))$, $\G$ be an environment and $\s$ a substitution
such that $\G\th l\s:list(A\s)$. We must prove that $\G\th
r\s:list(A\s)$. For $\G\th l\s:list(A\s)$, we must have a derivation
like~:

\begin{center}
{\small(symb)}~$\cfrac{
\G\th A'\s:\st \quad \G\th x\s:A'\s \quad \G\th \ell\s:list(A'\s)}
{\hs[-1cm]\mbox{\small(conv)}~\cfrac{
\begin{array}{c}
\G\th cons(A'\s,x\s,\ell\s):list(A'\s)\\
list(A'\s) \ad^* list(A\s) \quad \G\th list(A\s):\st\\
\end{array}}
{\hs[-1cm]\mbox{\small(symb)}~\cfrac{
\begin{array}{c}
\G\th cons(A'\s,x\s,\ell\s):list(A\s)\\
\G\th A\s:\st \quad \G\th \ell'\s:list(A\s)\\
\end{array}}
{\G\th l\s:list(A\s)}}}$
\end{center}

Therefore, $A'\s \ad^* A\s$ and we can derive $\G\th x\s:A\s$, $\G\th
\ell\s:list(A\s)$ and~:

\begin{center}
{\small(symb)}~$\cfrac{
\G\th A\s:\st \quad \G\th \ell\s:list(A\s) \quad \ell'\s:list(A\s)}
{\mbox{\small(symb)}~\cfrac{
\begin{array}{c}
\G\th app(A\s,\ell\s,\ell'\s):list(A\s)\\
\G\th A\s:\st \quad \G\th x\s:A\s\\
\end{array}}
{\G\th r\s:list(A\s)}}$
\end{center}

The point is that, although $l$ is not typable, from any typable
instance $l\s$ of $l$, we can deduce that $A'\s \ad^* A\s$. By this
way, we come to the following conditions~:

\begin{dfn}[Type-preserving rewrite rule]
\label{def-rew}
A rewrite rule $l\a r$ is {\em type-preserving\,} if there is an
environment $\G$ and a substitution $\r$ such that, if $l = f(\vl)$,
$\tf = \pvxT U$ and $\g = \vxl$ then~:

\begin{bfenumi}{S}
\item $dom(\r) \sle \FV(l) \moins \dom(\G)$,
\item $\G\th l\r:U\g\r$,
\item $\G\th r : U\g\r$,
\item for any substitution $\s$, environment $\D$ and type $T$, if
  $\D\th l\s:T$ then $\s:\G\a\D$,
\item for any substitution $\s$, environment $\D$ and type $T$, if
  $\D\th l\s:T$ then, for all $x\in \dom(\r)$, $x\s \ad^* x\r\s$.
\end{bfenumi}
\end{dfn}

In our example, it suffices to take $\G= A\!:\!\st, x\!:\!A,
\ell\!:\!list(A), \ell'\!:\!list(A)$ and $\r=\{A'\to A\}$.

One may wonder how to check these conditions. In practice, the symbols
are incrementally defined. So, assume that we have a confluent and
strongly normalizing CAC built over $\cF$ and $\cR$ and that we want
to add a new symbol $g$. Then, given $\G$ and $\r$, it is decidable to
check (S1) to (S3) in the CAC built over $\cF\cup\{g\}$ and $\cR$
since this system is confluent and strongly normalizing. In
\cite{blanqui01thesis}, we give a simple condition ensuring (S4) ($\G$
simply needs to be well chosen). The condition (S5) is the most
difficult to check and may require the confluence of $\a$.

%%% Local Variables: 
%%% mode: latex
%%% TeX-master: "main"
%%% End: 

%%%%%%%%%%%%%%%%%%%%%%%%%%%%%%%%%%%%%%%%%%%%%%%%%%%%%%%%%%%%%%%%%%%%%%%%%%%%%%
%%%%%%%%%%%%%%%%%%%%%%%%%%%%%%%%%%%%%%%%%%%%%%%%%%%%%%%%%%%%%%%%%%%%%%%%%%%%%%
%%%%%%%%%%%%%%%%%%%%%%%%%%%%%%%%%%%%%%%%%%%%%%%%%%%%%%%%%%%%%%%%%%%%%%%%%%%%%%

% admissibility conditions

%%%%%%%%%%%%%%%%%%%%%%%%%%%%%%%%%%%%%%%%%%%%%%%%%%%%%%%%%%%%%%%%%%%%%%%%%%%%%%
%%%%%%%%%%%%%%%%%%%%%%%%%%%%%%%%%%%%%%%%%%%%%%%%%%%%%%%%%%%%%%%%%%%%%%%%%%%%%%
%%%%%%%%%%%%%%%%%%%%%%%%%%%%%%%%%%%%%%%%%%%%%%%%%%%%%%%%%%%%%%%%%%%%%%%%%%%%%%

\section{Admissibility conditions}
\label{sec-admi}

%%%%%%%%%%%%%%%%%%%%%%%%%%%%%%%%%%%%%%%%%%%%%%%%%%%%%%%%%%%%%%%%%%%%%%%%%%%%%%
%%%%%%%%%%%%%%%%%%%%%%%%%%%%%%%%%%%%%%%%%%%%%%%%%%%%%%%%%%%%%%%%%%%%%%%%%%%%%%
%%%%%%%%%%%%%%%%%%%%%%%%%%%%%%%%%%%%%%%%%%%%%%%%%%%%%%%%%%%%%%%%%%%%%%%%%%%%%%

% inductive structure

%%%%%%%%%%%%%%%%%%%%%%%%%%%%%%%%%%%%%%%%%%%%%%%%%%%%%%%%%%%%%%%%%%%%%%%%%%%%%%
%%%%%%%%%%%%%%%%%%%%%%%%%%%%%%%%%%%%%%%%%%%%%%%%%%%%%%%%%%%%%%%%%%%%%%%%%%%%%%
%%%%%%%%%%%%%%%%%%%%%%%%%%%%%%%%%%%%%%%%%%%%%%%%%%%%%%%%%%%%%%%%%%%%%%%%%%%%%%

\subsection{Inductive structure}

Until now, we made few assumptions on symbols or rewrite rules. In
particular, we have no notion of inductive type. Yet, the structure of
inductive types plays a key role in strong normalization proofs
\cite{mendler87thesis}. On the other hand, we want rewriting to be as
general as possible by allowing matching on defined symbols and
equations among constructors. This is why, in the following, we
introduce an extended notion of constructor and a notion of inductive
structure which generalize usual definitions of inductive types
\cite{paulin93tlca}. Note that, in contrast with our previous work
\cite{blanqui99rta}, we allow inductive types to be polymorphic and
dependent, as it is the case in CIC.

%%%%%%%%%%%%%%%%%%%%%%%%%%%%%%%%%%%%%%%%%%%%%%%%%%%%%%%%%%%%%%%%%%%%%%%%%%%%%%
% constructors
%%%%%%%%%%%%%%%%%%%%%%%%%%%%%%%%%%%%%%%%%%%%%%%%%%%%%%%%%%%%%%%%%%%%%%%%%%%%%%

\newcommand{\Xs}{\cX^\st}
\newcommand{\XB}{\cX^\B}
\newcommand{\Fs}{\cF^\st}
\newcommand{\FB}{\cF^\B}
\newcommand{\FC}{\cC\FB}
\newcommand{\FD}{{\cD\FB}}
\newcommand{\co}{\cC o}

\newcommand{\ind}{\mi{Ind}}
\newcommand{\acc}{\mi{Acc}}

\begin{dfn}[Constructors]
  For $\cG \sle \cF$, let $\cR_\cG$ be the set of rules defining the
  symbols in $\cG$, that is, the rules whose left-hand side is headed
  by a symbol in $\cG$. The set of {\em free symbols\,} is $\cC\cF =
  \{ f \in \cF ~|~ \cR_{\{f\}} = \vide \}$. The set of {\em defined
    symbols\,} is $\cD\cF = \cF \moins \cC\cF$. The set of {\em
    constructors\,} of a free predicate symbol $C$ is $\co(C) = \{ f
  \in \Fs ~|~ \tf=\pvyU C(\vv) \mbox{ and } |\vy| = \at_f \}$.
\end{dfn}

The constructors of $C$ not only include the constructors in the usual
sense but every defined symbol whose output type is $C$. For example,
the symbols $0: int$, $s: int\a int$, $p: int\a int$, $+: int\a int\a
int$ and $\times: int\a int\a int$ defined by the rules $s(p(x)) \a
x$, $p(s(x)) \a x$ and others for $+$ and $\times$ are all
constructors of the type $int$ of integers.

%%%%%%%%%%%%%%%%%%%%%%%%%%%%%%%%%%%%%%%%%%%%%%%%%%%%%%%%%%%%%%%%%%%%%%%%%%%%%%
% inductive structure
%%%%%%%%%%%%%%%%%%%%%%%%%%%%%%%%%%%%%%%%%%%%%%%%%%%%%%%%%%%%%%%%%%%%%%%%%%%%%%

\begin{dfn}[Inductive structure]
An {\em inductive structure\,} is given by~:

\begin{lst}{\bu}
\item a quasi-ordering $\ge_\cF$ on $\cF$, called {\em precedence\,},
  whose strict part, $>_\cF$, is well-founded,
\item for each $C \in \FC$ such that $\tC = (\vx:\vT)\st$, a set
  $\ind(C) \sle \{ i \in \{ 1, .., \at_C \} ~|~ x_i \in \XB \}$ of
  {\em inductive\,} positions,
\item for each constructor $c$, a set $\acc(c) \sle \{ 1, .., \at_c
  \}$ of {\em accessible\,} positions.
\end{lst}
\end{dfn}

The accessible positions allow the user to describe which patterns can
be used for defining functions, and the inductive positions allow to
describe the arguments on which the free predicate symbols should be
monotone. This allows us to generalize the notion of positivity used
in CIC.

%%%%%%%%%%%%%%%%%%%%%%%%%%%%%%%%%%%%%%%%%%%%%%%%%%%%%%%%%%%%%%%%%%%%%%%%%%%%%%
% definition of positive and negative positions
%%%%%%%%%%%%%%%%%%%%%%%%%%%%%%%%%%%%%%%%%%%%%%%%%%%%%%%%%%%%%%%%%%%%%%%%%%%%%%

\newcommand{\pp}{\pos^+}
\renewcommand{\pm}{\pos^-}
\renewcommand{\pz}{\pos^0}
\newcommand{\pnz}{\pos^{\neq 0}}
\newcommand{\pd}{\pos^\d}

\begin{dfn}[Positive and negative positions]
  The sets of {\em positive\,} positions $\pp(T)$ and {\em negative\,}
  positions $\pm(T)$ of a term $T$ are mutually defined by induction
  on $T$ as follows~:

\begin{lst}{--}
\item $\pp(s) = \pp(F(\vt)) = \pp(X) = \{\vep\}$,
\item $\pm(s) = \pm(F(\vt)) = \pm(X) = \vide$,
\item $\pd(\px{V}W) = 1.\pos^{-\d}(V) \cup 2.\pd(W)$,
\item $\pd(\lx{V}W) = 1.\pos(V) \cup 2.\pd(W)$,
\item $\pd(Vu) = 1.\pd(V) \cup 2.\pos(u)$,
\item $\pd(VU) = 1.\pd(V)$,
\item $\pp(C(\vt)) = \{\vep\} \cup \bigcup \,\{ i.\pp(t_i) ~|~ i \in
  \ind(C) \}$,
\item $\pm(C(\vt)) = \bigcup \,\{ i.\pm(t_i) ~|~ i \in \ind(C) \}$,
\end{lst}

\noindent
where $\d \in \{ -, + \}$, $-+ = -$, $-- = +$.
\end{dfn}

For example, in $\px{A} B$, $B$ occurs positively while $A$ occurs
negatively. Now, with the type $list$ of polymorphic lists, $A$ occurs
positively in $list(A)$ iff $\ind(list) = \{1\}$.

%%%%%%%%%%%%%%%%%%%%%%%%%%%%%%%%%%%%%%%%%%%%%%%%%%%%%%%%%%%%%%%%%%%%%%%%%%%%%%
% inductive structure
%%%%%%%%%%%%%%%%%%%%%%%%%%%%%%%%%%%%%%%%%%%%%%%%%%%%%%%%%%%%%%%%%%%%%%%%%%%%%%

\begin{dfn}[Admissible inductive structure]
  An inductive structure is {\em admissible\,} if, for all $C \in \FC$
  with $\tC = \pvxT\st$~:

\begin{bfenumi}{I}
\item $\all i \in \ind(C)$, $v_i \in \XB$,
\end{bfenumi}

\noindent
and for all $c$ with $\tc = \pvyU C(\vv)$ and $j \in \acc(c)$~:

\begin{bfenumi}{I}
\addtocounter{enumi}{1}
\item $\all i \in \ind(C)$, $\pos(v_i,U_j) \sle \pp(U_j)$,
\item $\all D \IN \FC, D \!=_\cF\! C \!\A\! \pos(D,U_j) \!\sle\! \pp(U_j)$,
\item $\all D \in \FC, D >_\cF C \A \pos(D,U_j) = \vide$,
\item $\all F \in \FD, \pos(F,U_j) = \vide$,
\item $\all X \in \FVB(U_j), \ex\, \i_X \IN \{1, .., \at_C\}, v_{\i_X} = X$.
\end{bfenumi}
\end{dfn}

For example, with the type $list$ of polymorphic lists, $\ind(list)=
\{1\}$, $\acc(nil)= \{1\}$ and $\acc(cons)= \{1,2,3\}$ is an
admissible inductive structure. If we add the type $tree: \st$ and the
constructor $node: list(tree)\a tree$ with $\acc(node)= \{1\}$, we
still have an admissible structure.

The condition (I6) means that the predicate-arguments of a constructor
must be parameters of the type they define. One can find a similar
condition in the work of Walukiewicz \cite{walukiewicz00lfm} (called
``$\st$-dependency'') and in the work of Stefanova
\cite{stefanova98thesis} (called ``safeness'').

On the other hand, there is no such explicit restriction in CIC. But
the elimination scheme is typed in such a way that no very interesting
function can be defined on a type not satisfying (I6). For example,
consider the type of heterogeneous non-empty lists (we use the CIC
syntax here) $listh= Ind(X:\st)\{C_1|C_2\}$ where $C_1= \pA\st \px{A}
X$ and $C_2= \pA\st \px{A}$ $X\a X$. The typing rule for the non
dependent elimination schema (Nodep$_{\st,\st}$) is~:

\begin{center}
  $\cfrac{\G\th \ell:listh \quad \G\th Q:\st \quad \all i,\,
    \G\th f_i:C_i\{listh,Q\}}{\G\th Elim(\ell,Q)\{f_1|f_2\} : Q}$
\end{center}

\noindent
where $C_1\{listh,Q\}= \pA\st \px{A} Q$ and $C_2\{listh,Q\}= \pA\st
\px{A} listh\a Q\a Q$. Since $Q$, $f_1$ and $f_2$ must be typable in
$\G$, the result of $f_1$ and $f_2$ cannot depend on $A$ or on $x$.
This means that it is possible to compute the length of such a list
but not to use an element of the list.

%%%%%%%%%%%%%%%%%%%%%%%%%%%%%%%%%%%%%%%%%%%%%%%%%%%%%%%%%%%%%%%%%%%%%%%%%%%%%%
% primitive, basic and strictly positive predicates symbols
%%%%%%%%%%%%%%%%%%%%%%%%%%%%%%%%%%%%%%%%%%%%%%%%%%%%%%%%%%%%%%%%%%%%%%%%%%%%%%

\begin{dfn}{\bf (Primitive, basic and strictly positive predicates)}
  A free predicate symbol $C$ is~:
\begin{lst}{\bu}
\item {\em primitive\,} if, for all $D =_\cF C$, for all constructor
  $d$ of type $\td= \pvyU D(\vw)$ and for all $j\in \acc(d)$, $U_j$ is
  either of the form $E(\vt)$ with $E <_\cF D$ and $E$ basic, or of
  the form $E(\vt)$ with $E =_\cF D$.
\item {\em basic\,} if, for all $D =_\cF C$, for all constructor $d$
  of type $\td= \pvyU D(\vw)$ and for all $j\in \acc(d)$, if $E =_\cF
  D$ occurs in $U_j$ then $U_j$ is of the form $E(\vt)$.
\item {\em strictly positive\,} if, for all $D =_\cF C$, for all
  constructor $d$ of type $\td= \pvyU D(\vw)$ and for all $j\in
  \acc(d)$, if $E =_\cF D$ occurs in $U_j$ then $U_j$ is of the form
  $\pvzV E(\vt)$ and no occurrence of $D' =_\cF D$ occurs in $\vV$.
\end{lst}
\end{dfn}

For example, the type $list$ of polymorphic lists is basic but not
primitive. The type $listint$ of lists of integers with the
constructors $nilint:listint$ and $consint: int\a listint\a listint$
is primitive. And the type $ord$ of Brouwer's ordinals with the
constructors $0:ord$, $s:ord\a ord$ and $lim: (nat\a ord)\a ord$ is
strictly positive.

Although we do not explicitly forbid to have non-strictly positive
predicate symbols, the admissibility conditions we are going to
describe in the following subsections will not enable us to define
functions on such a predicate. The same restriction applies on CIC
while the system of Walukiewicz \cite{walukiewicz00lfm} is restricted
to basic predicates and the $\la R$-cube \cite{barbanera97jfp} or NDM
\cite{dowek98types} are restricted to primitive and non-dependent
predicates. However, in the following, for lack of space, we will
restrict our attention to basic predicates.

%%%%%%%%%%%%%%%%%%%%%%%%%%%%%%%%%%%%%%%%%%%%%%%%%%%%%%%%%%%%%%%%%%%%%%%%%%%%%%
%%%%%%%%%%%%%%%%%%%%%%%%%%%%%%%%%%%%%%%%%%%%%%%%%%%%%%%%%%%%%%%%%%%%%%%%%%%%%%
%%%%%%%%%%%%%%%%%%%%%%%%%%%%%%%%%%%%%%%%%%%%%%%%%%%%%%%%%%%%%%%%%%%%%%%%%%%%%%

% general schema

%%%%%%%%%%%%%%%%%%%%%%%%%%%%%%%%%%%%%%%%%%%%%%%%%%%%%%%%%%%%%%%%%%%%%%%%%%%%%%
%%%%%%%%%%%%%%%%%%%%%%%%%%%%%%%%%%%%%%%%%%%%%%%%%%%%%%%%%%%%%%%%%%%%%%%%%%%%%%
%%%%%%%%%%%%%%%%%%%%%%%%%%%%%%%%%%%%%%%%%%%%%%%%%%%%%%%%%%%%%%%%%%%%%%%%%%%%%%

\subsection{General Schema}

\renewcommand{\sf}{_{stat_f}}
\newcommand{\sF}{_{stat_F}}
\newcommand{\sg}{_{stat_g}}

\newcommand{\stat}{\mr{stat}}
\newcommand{\as}{\alpha^\stat}

\newcommand{\CC}{\mr{CC}}

The constructors of primitive predicates (remember that they include
all symbols whose output type is a primitive predicate), defined by
usual first-order rules, are easily shown to be strongly normalizing
since the combination of first-order rewriting with $\ab$ preserves
strong normalization \cite{breazu91tcs}.

\vs[2mm]
On the other hand, in the presence of higher-order rules, few
techniques are known~:

\begin{lst}{\bu}
  
\item Van de Pol \cite{vandepol96thesis} extended to the higher-order
  case the use of strictly monotone interpretations . This technique
  is very powerful but difficult to use in practice and has not been
  studied yet in type systems richer than the simply-typed
  $\la$-calculus.
  
\item Jouannaud and Okada \cite{jouannaud97tcs} defined a syntactic
  criterion, the General Schema, which extends primitive recursive
  definitions. This schema has been reformulated and enhanced to deal
  with definitions on strictly-positive types \cite{blanqui01tcs}, to
  higher-order pattern-matching \cite{blanqui00rta} and to richer type
  systems with object-level rewriting
  \cite{barbanera97jfp,blanqui99rta}.
  
\item Jouannaud and Rubio \cite{jouannaud99lics} extended to the
  higher-order case the use of Dershowitz's recursive path ordering.
  The obtained ordering can be seen as a recursive version of the
  General Schema and has been extended by Walukiewicz
  \cite{walukiewicz00lfm} to the Calculus of Constructions with
  object-level rewriting.

\end{lst}

\vs[2mm]
Here, we present an extension of the General Schema defined in
\cite{blanqui99rta} to deal with type-level rewriting, the main
novelty of our paper.

The General Schema is based on Tait and Girard's computability
predicate technique \cite{girard88book} for proving the strong
normalization of the simply-typed $\la$-calculus and system F. This
technique consists in interpreting each type $T$ by a set $\I{T}$ of
strongly normalizable terms, called {\em computable\,}, and in proving
that $t \in \I{T}$ whenever $\G \th t:T$.

The idea of the General Schema is then to define, from a left-hand
side of rule $f(\vl)$, a set of right-hand sides $r$ that are
computable whenever the $l_i$'s are computable. This set is built from
the variables of the left-hand side, called {\em accessible\,}, that
are computable whenever the $l_i$'s are computable, and is then closed
by computability-preserving operations.

\vs[2mm] For the sake of simplicity, two sequences of arguments of a
symbol $f$ will be compared in a lexicographic manner. But it is
possible to do these comparisons in a multiset manner or with a simple
combination of lexicographic and multiset comparisons (see
\cite{blanqui01thesis} for details).

%%%%%%%%%%%%%%%%%%%%%%%%%%%%%%%%%%%%%%%%%%%%%%%%%%%%%%%%%%%%%%%%%%%%%%%%%%%%%%
% accessibility relation
%%%%%%%%%%%%%%%%%%%%%%%%%%%%%%%%%%%%%%%%%%%%%%%%%%%%%%%%%%%%%%%%%%%%%%%%%%%%%%

\begin{dfn}[Accessibility]
  A pair $\ps{u,U}$ is {\em accessible\,} in a pair $\ps{t,T}$,
  written $\ps{t,T} \tgt_1 \ps{u,U}$, if $\ps{t,T} =
  \ps{c(\vu),C(\vv)\g}$ and $\ps{u,U} = \ps{u_j,U_j\g}$ with $c$ a
  constructor of type $\tc=\pvyU C(\vv)$, $\g=\vyu$ and $j\in
  \acc(c)$.
\end{dfn}

For example, in the definition of $app$ previously given, $A'$, $x$
and $\ell$ are all accessible in $t=cons(A',x,\ell)$~: $\ps{t,list(A)}
\tgt_1 \ps{A',\st}$, $\ps{t,list(A)} \tgt_1 \ps{x,A'}$ and
$\ps{t,list(A)} \tgt_1 \ps{\ell,list(A')}$.

%%%%%%%%%%%%%%%%%%%%%%%%%%%%%%%%%%%%%%%%%%%%%%%%%%%%%%%%%%%%%%%%%%%%%%%%%%%%%%
% derived type
%%%%%%%%%%%%%%%%%%%%%%%%%%%%%%%%%%%%%%%%%%%%%%%%%%%%%%%%%%%%%%%%%%%%%%%%%%%%%%

\begin{dfn}[Derived type]
  Let $t$ be a term of the form $l\s$ with $l=f(\vl)$ algebraic,
  $\tf=\pvxT U$ and $\g=\vxl$.  Let $p\in \pos(l)$ with $p\neq\vep$.
  The subterm $t|_p$ of $t$ has a {\em derived type\,}, $\tau(t,p)$,
  defined as follows~:

\begin{lst}{--}
\item if $p=i$ then $\tau(t,p) = T_i\g\s$,
\item if $p=iq$ and $q\neq \vep$ then $\tau(t,p) = \tau(t_i,q)$.
\end{lst}
\end{dfn}

%%%%%%%%%%%%%%%%%%%%%%%%%%%%%%%%%%%%%%%%%%%%%%%%%%%%%%%%%%%%%%%%%%%%%%%%%%%%%%
% well-formed rule
%%%%%%%%%%%%%%%%%%%%%%%%%%%%%%%%%%%%%%%%%%%%%%%%%%%%%%%%%%%%%%%%%%%%%%%%%%%%%%

\begin{dfn}[Well-formed rule]
  Let $R= (l\a r,$ $\G,\r)$ be a rule with $l=f(\vl)$, $\tf= \pvxT U$
  and $\g=\vxl$. The rule $R$ is {\em well-formed\,} if, for all $x\in
  \dom(\G)$, there is $i\le \at_f$ and $p_x\in \pos(x,l_i)$ such that
  $\ps{l_i,T_i\g} \,\tgt_1^*\, \ps{x,\tau(l,ip_x)}$ and
  $\tau(l,ip_x)\r = x\G$.
\end{dfn}

%%%%%%%%%%%%%%%%%%%%%%%%%%%%%%%%%%%%%%%%%%%%%%%%%%%%%%%%%%%%%%%%%%%%%%%%%%%%%%
% computable closure
%%%%%%%%%%%%%%%%%%%%%%%%%%%%%%%%%%%%%%%%%%%%%%%%%%%%%%%%%%%%%%%%%%%%%%%%%%%%%%

\newcommand{\thc}{\th_\mr{\!\!c}}

\begin{dfn}[Computable closure]
  Let $R= (l\a r,\G_0,\r)$ be a rule with $l=f(\vl)$, $\tf= \pvxT U$
  et $\g=\vxl$. The order $>$ on the arguments of $f$ is the
  lexicographic extension of $\tgt_1^+$. The {\em computable
    closure\,} of $R$ is the relation $\thc$ defined by the rules of
  Figure~\ref{fig-comp-closure}.
\end{dfn}

%%%%%%%%%%%%%%%%%%%%%%%%%%%%%%%%%%%%%%%%%%%%%%%%%%%%%%%%%%%%%%%%%%%%%%%%%%%%%%
% rules of the computable closure
%%%%%%%%%%%%%%%%%%%%%%%%%%%%%%%%%%%%%%%%%%%%%%%%%%%%%%%%%%%%%%%%%%%%%%%%%%%%%%

\begin{figure}[ht]
\begin{center}
\caption{Computable closure\label{fig-comp-closure}}
\vs[4mm]
\begin{tabular}{r@{~}c}
{\small(acc)} & $\cfrac{\G_0\thc x\G_0:s \quad x\in\dom^s(\G_0)}
{\G_0\thc x:x\G_0}$\\

{\small(ax)} & $\cfrac{}{\G_0\thc \st : \B}$\\

\\{\small(symb$^<$)} & $\cfrac{
\begin{array}{c}
g\in \cF^s_n,\, \tg = \pvyU V,\, \g = \vyu\\
g <_\cF f \quad \G\thc \tg:s
\quad \all i,\, \G\thc u_i:U_i\g\\
\end{array}
}{\G \thc g(\vu) : V\g}$\\

\\{\small(symb$^=$)} & $\cfrac{
\begin{array}{c}
g\in \cF^s_n,\, \tg = \pvyU V,\, \g = \vyu\\
g =_\cF f \quad \G\thc \tg:s
\quad \all i,\, \G\thc u_i:U_i\g\\
\ps{\vl,\vT\g_0} > \ps{\vu,\vU\g}\\
\end{array}
}{\G \thc g(\vu) : V\g}$\\

\\{\small(var)} & $\cfrac{\G\thc T:s \quad x\in\cX^s\moins\FV(l)}
{\GxT\thc x:T}$\\

\\{\small(weak)} & $\cfrac{\G\thc t:T \quad \G\thc U:s \quad
x\in\cX^s\moins\FV(l)}{\GxU \thc t:T}$\\

\\{\small(prod)} & $\cfrac{\G\thc T:s \quad \GxT\thc U:s'}
{\G\thc \px{T}U:s'}$\\

\\{\small(abs)} & $\cfrac{\GxT \thc u:U \quad \G\thc \px{T}U:s}
{\G\thc \lx{T}u : \px{T}U}$\\

\\{\small(app)} & $\cfrac{\G\thc t:\px{U}V \quad \G\thc u:U}
{\G\thc tu:V\xu}$ \\

\\{\small(conv)} & $\cfrac{\G\thc t:T \quad T \ad^* T' \quad
\G\thc T':s'}{\G\thc t:T'}$\\
\end{tabular}
\end{center}
\end{figure}

%%%%%%%%%%%%%%%%%%%%%%%%%%%%%%%%%%%%%%%%%%%%%%%%%%%%%%%%%%%%%%%%%%%%%%%%%%%%%%
% general schema
%%%%%%%%%%%%%%%%%%%%%%%%%%%%%%%%%%%%%%%%%%%%%%%%%%%%%%%%%%%%%%%%%%%%%%%%%%%%%%

\begin{dfn}[General Schema]
  A rule $(f(\vl)\a r,\G,\r)$ with $\tf= \pvxT U$ and $\g=\vxl$
  satisfies the {\em General Schema} if it is well-formed and $\G\thc
  r:U\g\r$.
\end{dfn}

It is easy to check that the rules for $app$ are well-formed and that
$\G\thc cons(A,x,app(A,\ell,\ell')):list(A)$. For example, we show
that $\G\thc app(A,\ell,\ell'):list(A)$~:

\begin{center}
$\cfrac{
\begin{array}{c}
\cfrac{\G\thc \st:\B}{\G\thc A:\st}
\quad
\cfrac{\cfrac{\cfrac{\ldots}{\G\thc A:\st}}{\G\thc
    list(A):\st}}{\G\thc \ell:list(A)}
\quad
\cfrac{\ldots}{\G\thc \ell':list(A)}\\
\ps{cons(A',x,\ell),list(A)} > \ps{\ell,list(A)}\\
\end{array}}
{\G\thc app(A,\ell,\ell')}$
\end{center}

%%%%%%%%%%%%%%%%%%%%%%%%%%%%%%%%%%%%%%%%%%%%%%%%%%%%%%%%%%%%%%%%%%%%%%%%%%%%%%
%%%%%%%%%%%%%%%%%%%%%%%%%%%%%%%%%%%%%%%%%%%%%%%%%%%%%%%%%%%%%%%%%%%%%%%%%%%%%%
%%%%%%%%%%%%%%%%%%%%%%%%%%%%%%%%%%%%%%%%%%%%%%%%%%%%%%%%%%%%%%%%%%%%%%%%%%%%%%

% admissibility conditions

%%%%%%%%%%%%%%%%%%%%%%%%%%%%%%%%%%%%%%%%%%%%%%%%%%%%%%%%%%%%%%%%%%%%%%%%%%%%%%
%%%%%%%%%%%%%%%%%%%%%%%%%%%%%%%%%%%%%%%%%%%%%%%%%%%%%%%%%%%%%%%%%%%%%%%%%%%%%%
%%%%%%%%%%%%%%%%%%%%%%%%%%%%%%%%%%%%%%%%%%%%%%%%%%%%%%%%%%%%%%%%%%%%%%%%%%%%%%

\subsection{Admissibility conditions}

%%%%%%%%%%%%%%%%%%%%%%%%%%%%%%%%%%%%%%%%%%%%%%%%%%%%%%%%%%%%%%%%%%%%%%%%%%%%%%
% rewrite systems
%%%%%%%%%%%%%%%%%%%%%%%%%%%%%%%%%%%%%%%%%%%%%%%%%%%%%%%%%%%%%%%%%%%%%%%%%%%%%%

\newcommand{\domB}{\dom^\B}

\begin{dfn}[Rewrite systems]
  Let $\cG$ be a set of symbols. The {\em rewrite system\,}
  $(\cG,\cR_\cG)$ is~:

\begin{lst}{\bu}
\item {\em algebraic\,} if~:
\begin{lst}{--}
\item $\cG$ is made of predicate symbols or of constructors of
  primitive predicates,
\item all rules of $\cR_\cG$ have an algebraic right-hand side;
\end{lst}

\item {\em non-duplicating\,} if, for all $l\a r \in \cR_\cG$, no
  variable has more occurrences in $r$ than in $l$;
  
\item {\em primitive\,} if, for all rule $l\a r \in \cR_\cG$, $r$ is
  of the form $[\vx:\vT]g(\vu)\vv$ with $g$ belonging to $\cG$ or
  $g$ being a primitive predicate symbol;

\item {\em simple\,} if, for all $g(\vl)\a r \in \cR_\cG$~:
\begin{lst}{--}
\item all the symbols occuring in $\vl$ are free,
\item for all sequence of terms $\vt$, at most one rule can apply at
  the top of $g(\vt)$,
\item for all rule $g(\vl)\a r \in \cR_\cG$ and all $Y\in \FVB(r)$,
  there is a unique $\k_Y$ such that $l_{\k_Y}=Y$;
\end{lst}

\item {\em positive\,} if, for all $l\a r \in \cR_\cG$ and all
  $g\in\cG$, $\pos(g,r) \sle \pp(r)$;
  
\item {\em recursive\,} if all the rules of $\cR_\cG$ satisfy the
  General Schema;
  
\item {\em safe\,} if, for all $(g(\vl)\!\a\! r,\G,\r)\in \cR_\cG$
  with $\tg= \pvxT$ $U$ and $\g=\vxl$~:
\begin{lst}{--}
\item for all $X\in \FVB(\vT U)$, $X\g\r\in \domB(\G)$,
\item for all $X,X'\!\in\! \FVB(\vT U)$, $X\g\r\!=\!X'\g\r \A
  X\!=\!X'$.
\end{lst}
\end{lst}
\end{dfn}

%%%%%%%%%%%%%%%%%%%%%%%%%%%%%%%%%%%%%%%%%%%%%%%%%%%%%%%%%%%%%%%%%%%%%%%%%%%%%%
% admissible CAC
%%%%%%%%%%%%%%%%%%%%%%%%%%%%%%%%%%%%%%%%%%%%%%%%%%%%%%%%%%%%%%%%%%%%%%%%%%%%%%

\newcommand{\na}{_{na}}

\begin{dfn}[Admissible CAC]
  A CAC is {\em admissible\,} if~:

\begin{bfenumi}{A}
\item $\a = \ar \cup \ab$ is confluent;
  
\item its inductive structure is admissible;

\item $(\FD,\cR_\FD)$ is either~:
\begin{lst}{--}
\item primitive,
\item simple and positive,
\item simple and recursive;
\end{lst}
  
\item there is a partition $\cF_a \uplus \cF\na$ of $\cD\cF$ ({\em
    algebraic\,} and {\em non-algebraic\,} symbols) such that~:
\begin{lst}{--}
\item $(\cF_a,\cR_{\cF_a})$ is algebraic, non-duplicating and strongly
  normalizing,
\item no symbol of $\cF\na$ occurs in the rules of $\cR_{\cF_a}$,
\item $(\cF\na,\cR_{\cF\na})$ is safe and recursive.
\end{lst}
\end{bfenumi}
\end{dfn}

The simplicity condition in (A3) extends to the case of rewriting the
restriction in CIC of strong elimination to ``small'' inductive types,
that is, to the types whose constructors have no predicate-arguments
except the parameters of the type.

The safeness condition in (A4) means that one cannot do
pattern-matching or equality tests on predicate-arguments that are
necessary for typing other arguments. In her extension of HORPO to the
Calculus of Constructions, Walukiewicz requires similar conditions
\cite{walukiewicz00lfm}.

The non-duplication condition in (A4) ensures the modularity of the
strong normalization. Indeed, in general, the combination of two
strongly normalizing rewrite systems is not strongly normalizing.

Now, for proving (A1), one can use the following result of van Oostrom
\cite{oostrom94thesis} (remember that $\cR\cup\b$ can be seen as a CRS
\cite{klop93tcs})~: the combination of two confluent left-linear CRS's
having no critical pairs between each other is confluent. So, since
$\ab$ is confluent and $\cR$ and $\b$ cannot have critical pairs
between each other, if $\cR$ is left-linear and confluent then
$\ar\cup\ab$ is confluent. Therefore, our conditions (S1) to (S5) are
very useful to eliminate the non-linearities due to typing reasons.

%%%%%%%%%%%%%%%%%%%%%%%%%%%%%%%%%%%%%%%%%%%%%%%%%%%%%%%%%%%%%%%%%%%%%%%%%%%%%%
% strong normalization
%%%%%%%%%%%%%%%%%%%%%%%%%%%%%%%%%%%%%%%%%%%%%%%%%%%%%%%%%%%%%%%%%%%%%%%%%%%%%%

We can now state our main result. You can find a detailed proof in
\cite{blanqui01thesis}.

\begin{thm}[Strong normalization]
  Any admissible CAC is strongly normalizing.
\end{thm}

The proof is based on Coquand and Gallier's extension to the Calculus
of Constructions \cite{coquand90lf} of Tait and Girard's computability
predicate technique \cite{girard88book}. As explained before, the idea
is to define an interpretation for each type and to prove that each
well-typed term belongs to the interpretation of its type.

The main difficulty is to define an interpretation for predicate
symbols that is invariant by reduction, a condition required by the
type conversion rule (conv).

Thanks to the positivity conditions, the interpretation of a free
predicate symbol can be defined as the least fixpoint of a monotone
function over the lattice of computability predicates.

For the defined predicate symbols, it depends on the kind of system
$(\FD,\cR_\FD)$ is. If it is primitive then we simply interpret it as
the set of strongly normalizable terms. If it is positive then, thanks
to the positivity condition, we can interpret it as a least fixpoint.
Finally, if it is recursive then we can define its interpretation
recursively, the General Schema providing a well-founded definition.

%%% Local Variables: 
%%% mode: latex
%%% TeX-master: "main"
%%% End: 

%%%%%%%%%%%%%%%%%%%%%%%%%%%%%%%%%%%%%%%%%%%%%%%%%%%%%%%%%%%%%%%%%%%%%%%%%%%%%%
%%%%%%%%%%%%%%%%%%%%%%%%%%%%%%%%%%%%%%%%%%%%%%%%%%%%%%%%%%%%%%%%%%%%%%%%%%%%%%
%%%%%%%%%%%%%%%%%%%%%%%%%%%%%%%%%%%%%%%%%%%%%%%%%%%%%%%%%%%%%%%%%%%%%%%%%%%%%%

% examples

%%%%%%%%%%%%%%%%%%%%%%%%%%%%%%%%%%%%%%%%%%%%%%%%%%%%%%%%%%%%%%%%%%%%%%%%%%%%%%
%%%%%%%%%%%%%%%%%%%%%%%%%%%%%%%%%%%%%%%%%%%%%%%%%%%%%%%%%%%%%%%%%%%%%%%%%%%%%%
%%%%%%%%%%%%%%%%%%%%%%%%%%%%%%%%%%%%%%%%%%%%%%%%%%%%%%%%%%%%%%%%%%%%%%%%%%%%%%

\section{Examples}
\label{sec-examples}

%%%%%%%%%%%%%%%%%%%%%%%%%%%%%%%%%%%%%%%%%%%%%%%%%%%%%%%%%%%%%%%%%%%%%%%%%%%%%%
% CIC
%%%%%%%%%%%%%%%%%%%%%%%%%%%%%%%%%%%%%%%%%%%%%%%%%%%%%%%%%%%%%%%%%%%%%%%%%%%%%%

\subsection{Calculus of Inductive Constructions}

\newcommand{\vA}{\vec{A}}
\newcommand{\vB}{\vec{B}}
\newcommand{\vC}{\vec{C}}
\newcommand{\vf}{\vec{f}}
\newcommand{\vb}{\vec{b}}
\newcommand{\vm}{\vec{m}}
\newcommand{\pvxA}{(\vx:\vA)}
\newcommand{\pvzB}{(\vz:\vB)}

We are going to see that we can apply our strong normalization theorem
to a sub-system of CIC \cite{paulin93tlca} by translating it into an
admissible CAC. The first complete proof of strong normalization of
CIC (with strong elimination) is due to Werner \cite{werner94thesis}
who, in addition, considers $\eta$-reductions in the type conversion
rule.

\vs[2mm]
In CIC, one has strictly-positive inductive types and the
corresponding induction principles. We recall the syntax and the
typing rules of CIC but, for the sake of simplicity, we will restrict
our attention to basic inductive types and non-dependent elimination
schemas. For a complete presentation, see \cite{blanqui01thesis}.

\begin{lst}{\bu}
\item Inductive types are denoted by $Ind\p{X}{A}\{\vC\}$ where the
  $C_i$'s are the types of the constructors. The term $A$ must be of
  the form $\pvxA\st$ and the $C_i$'s of the form $(\vz:\vB)X\vm$.
\item The $i$-th constructor of an inductive type $I$ is denoted by
  $Constr(i,I)$.
\item Recursors are denoted by $Elim(I,Q,\va,c)$ where $I$ is the
  inductive type, $Q$ the type of the result, $\va$ the arguments of
  $I$ and $c$ a term of type $I\va$.
\end{lst}

\vs[2mm]
The typing rules for these constructions are given in
Figure~\ref{fig-CIC-typing-rules}. The rules for the other
constructions are the same as for the Calculus of Constructions.

\begin{figure}[ht]
\begin{center}
%\vs[2mm]
\caption{Typing rules of CIC\label{fig-CIC-typing-rules}}
\begin{tabular}{r@{~}c}
\\{\small(Ind$_\st$)} & $\cfrac{\all i,\, \G,X:A\th C_i:\st}
{\G\th Ind\pX{A}\{\vC\}:A}$\\

\\{\small(Constr)} & $\cfrac{\G\th I=Ind\pX{A}\{\vC\}:A}
{\G\th Constr(i,I):C_i\{X\to I\}}$\\

\\{\small(Nodep$_{\st,s}$)} & $\cfrac{
\begin{array}{c}
\G\th c:I\va \quad \G\th Q:\pvxA s\\
\all i,\, \G\th f_i:C_i\{I,Q\}\\
\end{array}}
{\G\th Elim(I,Q,\va,c)\{\vf\}:Q\va}$\\
\end{tabular}
\end{center}
\vs[-5mm]
\end{figure}

If $C_i\!=\! \p{\vz}{\vB} X\vm$ then $C_i\{I,Q\}$ denotes
$\p{\vz}{\vB} \p{\vz'}{\vB\{X\to Q\}}\, Q\vm$. The reduction relation
associated to $Elim$ is called {\em $\i$-reduction\,} and is defined
as follows~:

\begin{center}
  $Elim(I,Q,\va,Constr(i,I')\,\vb)\{\vf\} \a_\i f_i\, \vb~ \vb'$
\end{center}

\noindent
where, if $C_i=\pvzB X\vm$, then $b_j'= Elim(I,Q,\va',b_j)$ if
$B_j=X\va'$, and $b_j'=b_j$ otherwise.

\vs[2mm]
Now, we consider the sub-system CIC$^-$ obtained by applying the
following restrictions~:

\begin{lst}{\bu}
\item In the typing rules (Ind$_\st$) and (Constr), we assume that
  $\G$ is empty since, in CAC, the types of the symbols must be
  typable in the empty environment.
\item In the rule (Nodep$_{\st,\st}$) (the one for weak elimination),
  we require $Q$ to be typable in the empty environment.
\item In the rule (Nodep$_{\st,\B}$) (the one for strong elimination),
  instead of requiring $\G\th Q:\pvxA\B$ which is not possible in the
  Calculus of Constructions since $\B$ is not typable, we require $Q$
  to be a closed term of the form $[\vx:\vA]K$ with $K$ of the form
  $\pvyU\st$.
\item We assume that every inductive type satisfies (I6).
\end{lst}

\begin{thm}
  CIC$^-$ can be translated into an admissible CAC, hence is strongly
  normalizing.
\end{thm}

We define the translation $\ps{~}$ by induction on the size of terms~:

\begin{lst}{\bu}
\item Let $I=Ind\pX{A}\{\vC\}$. We define $\ps{I}= [\vx:\ps{\vA}]$
  $Ind_I(\vx)$ where $Ind_I$ is a symbol of type
  $\p{\vx}{\ps{\vA}}\st$.
\item By assumption, $C_i= \p{\vz}{\vB} X\vm$. We define
  $\ps{Constr(i,I)}= \l{\vz}{\vB} Constr_I^i(\vz)$ where $Constr_I^i$
  is a symbol of type $(\vz:\ps{\vB}) Ind_I(\ps{\vm})$.
\item Let $T_i= C_i\{I,Q\}$. If $Q= \l{\vx}{\vA} K$ then we define
  $\ps{Elim(I,Q,\va,c)\{\vf\}}= SElim_I^Q(\ps{\vf},\ps{\va},\ps{c})$
  where $SElim_I^Q$ is a symbol of type $\p{\vf}{\ps{\vT}}$
  $\p{\vx}{\ps{\vA}}$ $\ps{K}$. Otherwise, we define
  $\ps{Elim(I,Q,\va,c)\{\vf\}}=
  W\!Elim_I(\ps{Q},\ps{\vf},\ps{\va},\ps{c})$ where $W\!Elim_I$ is a
  symbol of type $\p{Q}{\ps{A}} \p{\vf}{\ps{\vT}} \p{\vx}{\ps{\vA}}\,
  \ps{Q}\vx$.
\item The other terms are defined recursively ($\ps{uv}= \ps{u}\ps{v},
  \ldots$).
\end{lst}

The $\i$-reduction is translated by the following rules~:

\begin{rewc}
  SElim_I^Q(\vf,\va,Constr_I^i(\vb)) & f_i\, \vb~ \vb'\\
  W\!Elim_I(Q,\vf,\va,Constr_I^i(\vb)) & f_i\, \vb~ \vb'\\
\end{rewc}

\noindent
where, if $C_i=\pvzB X\vm$, then $b_j'= SElim_I^Q(\vf,\va',b_j)$ (or
$W\!Elim_I(Q,\vf,\va',b_j)$) if $B_j=X\va'$, and $b_j'=b_j$ otherwise.

\vs[2mm]
Now, we are left to check the admissibility~:

\begin{enumi}{A}
\item $\a_{\b\i}$ is orthogonal, hence confluent
  \cite{oostrom94thesis}.
  
\item The inductive structure defined by $I <_\cF J$ if $I$ is a
  subterm of $J$, $\ind(Ind_I)=\vide$, $\acc(Constr_I^i)=
  \{1,..,|\vz|\}$ if $C_i=\pvzB X\vm$, is admissible.
  
\item The rules defining the strong recursors form a simple (they are
  defined by case on each constructor and only for small inductive
  types) and recursive rewrite system (they satisfy the General
  Schema).
  
\item The rules defining the recursors form a safe (except for the
  constructor, all the arguments are distinct variables) and recursive
  rewrite system (they satisfy the General Schema).
\end{enumi}

%%%%%%%%%%%%%%%%%%%%%%%%%%%%%%%%%%%%%%%%%%%%%%%%%%%%%%%%%%%%%%%%%%%%%%%%%%%%%%
% NDM
%%%%%%%%%%%%%%%%%%%%%%%%%%%%%%%%%%%%%%%%%%%%%%%%%%%%%%%%%%%%%%%%%%%%%%%%%%%%%%

\subsection{Natural Deduction Modulo}

NDM for first-order logic \cite{dowek98trtpm} can be presented as an
extension of Natural Deduction with the additional inference rule~:

\begin{center}
$\cfrac{\G\th P}{\G\th Q}$ ~~if $P \equiv Q$
\end{center}

\noindent
where $\equiv$ is a congruence relation on propositions. This is a
powerful extension of first-order logic since both higher-order logic
and set theory with a comprehension symbol can be described in this
framework (by using explicit substitutions).

In \cite{dowek98types}, Dowek and Werner study the termination of
cut-elimination in the case where $\equiv$ is induced by a confluent
and weakly-normalizing rewrite system. In particular, they prove the
termination in two general cases~: when the rewrite system is positive
and when it is quantifier-free. In \cite{dowek00note}, they provide an
example of confluent and weakly normalizing rewrite system for which
cut-elimination is not terminating. The problem comes from the fact
that the elimination rule for $\all$ introduces a substitution~:

\begin{center}
$\cfrac{\G\th \all x.P(x)}{\G\th P(t)}$
\end{center}

Thus, when a predicate symbol is defined by a rule whose right-hand
side contains quantifiers, its combination with $\b$ may not preserve
normalization. Therefore, a criterion for higher-order rewriting is
needed.

\vs[2mm] Since NDM is a CAC (we can define the logical connectors as
inductive types), we can compare in more details the conditions of
\cite{dowek98types} with our conditions.

\begin{enumi}{A}
\item In \cite{dowek98types}, only $\ar$ is required to be confluent.
  In general, this is not sufficient for having the confluence of $\ar
  \cup \ab$. However, if $\cR$ is left-linear then $\ar \cup \ab$ is
  confluent \cite{oostrom94thesis}.

\item NDM types are primitive and form an admissible inductive
  structure if we take them equivalent in the relation $\le_\cF$.
  
\item In \cite{dowek98types}, the termination of cut-elimination is
  proved in two general cases~: when $(\FD,\cR_\FD)$ is
  quantifier-free and when it is positive. Quantifier-free rewrite
  systems are primitive. So, in this case, (A3) is satisfied. In the
  positive case, we require that left-hand sides are made of free
  symbols and that at most one rule can apply at the top of a term. On
  the other hand, we provide a new case~: $(\FD,\cR_\FD)$ can be
  simple and recursive.
  
\item Quantifier-free rules are algebraic and rules with quantifiers
  are not. In \cite{dowek98types}, these two kinds of rules are
  treated in the same way but the counter-example given in
  \cite{dowek00note} shows that they should not. In CAC, we require
  that the rules with quantifiers satisfy the General Schema.
\end{enumi}

\begin{thm}
  A NDM system satisfying (A1), (A3) and (A4) is admissible, hence
  strongly normalizing.
\end{thm}

%%%%%%%%%%%%%%%%%%%%%%%%%%%%%%%%%%%%%%%%%%%%%%%%%%%%%%%%%%%%%%%%%%%%%%%%%%%%%%
% CIC + R
%%%%%%%%%%%%%%%%%%%%%%%%%%%%%%%%%%%%%%%%%%%%%%%%%%%%%%%%%%%%%%%%%%%%%%%%%%%%%%

\subsection{CIC + Rewriting}

As a combination of the two previous applications, our work shows that
the extension of CIC$^-$ with user-defined rewrite rules, even at the
predicate-level, is sound if these rules follow our admissibility
conditions.

As an example, we consider simplification rules on propositions that
are not definable in CIC. Assume that we have the symbols
$\ou\!:\!\st\a\st\a\st$, $\et\!:\!\st\a\st\a\st$, $\non:\st\a\st$,
$\bot:\st$, $\top:\st$, and the rules~:

\begin{center}
\begin{tabular}{ccc}
\begin{rew}
\top\ou P & \top\\
P\ou\top & \top\\
\end{rew}
&
\begin{rew}
\bot\et P & \bot\\
P\et\bot & \bot\\
\end{rew}
&
\begin{rew}
\non \top & \bot\\
\non \bot & \top\\
\end{rew}
\end{tabular}\\[2mm]

$\non(P\et Q) \a \non P \ou \non Q
\quad\quad \non(P\ou Q) \a \non P \et \non Q$
\end{center}

\noindent
The predicate constructors $\ou$, $\et$, \ldots are all primitive. The
rewrite system is primitive, algebraic, strongly normalizing and
confluent (this can be automatically proved by CiME \cite{cime}).
Since it is left-linear, its combination with $\ab$ is confluent
\cite{oostrom94thesis}. Therefore, it is an admissible CAC. But it
lacks many other rules \cite{hsiang85ai} which requires rewriting
modulo associativity and commutativity, an extension we leave for
future work.

%%% Local Variables: 
%%% mode: latex
%%% TeX-master: "main"
%%% End: 

%%%%%%%%%%%%%%%%%%%%%%%%%%%%%%%%%%%%%%%%%%%%%%%%%%%%%%%%%%%%%%%%%%%%%%%%%%%%%%
%%%%%%%%%%%%%%%%%%%%%%%%%%%%%%%%%%%%%%%%%%%%%%%%%%%%%%%%%%%%%%%%%%%%%%%%%%%%%%
%%%%%%%%%%%%%%%%%%%%%%%%%%%%%%%%%%%%%%%%%%%%%%%%%%%%%%%%%%%%%%%%%%%%%%%%%%%%%%

% conclusion

%%%%%%%%%%%%%%%%%%%%%%%%%%%%%%%%%%%%%%%%%%%%%%%%%%%%%%%%%%%%%%%%%%%%%%%%%%%%%%
%%%%%%%%%%%%%%%%%%%%%%%%%%%%%%%%%%%%%%%%%%%%%%%%%%%%%%%%%%%%%%%%%%%%%%%%%%%%%%
%%%%%%%%%%%%%%%%%%%%%%%%%%%%%%%%%%%%%%%%%%%%%%%%%%%%%%%%%%%%%%%%%%%%%%%%%%%%%%

\section{Conclusion}
\label{sec-conclu}

We have defined an extension of the Calculus of Constructions by
functions and predicates defined with rewrite rules. The main
contributions of our work are the following~:

\begin{lst}{\bu}
\item We consider a general notion of rewriting at the predicate-level
  which generalizes the ``strong elimination'' of the Calculus of
  Inductive Constructions \cite{paulin93tlca,werner94thesis}. For
  example, we can define simplification rules on propositions that are
  not definable in CIC.
  
\item We consider general syntactic conditions, including confluence,
  that ensure the strong normalization of the calculus. In particular,
  these conditions are fulfilled by two important systems~: a
  sub-system of the Calculus of Inductive Constructions which is the
  basis of the proof assistant Coq \cite{coq}, and the Natural
  Deduction Modulo \cite{dowek98trtpm,dowek98types} a large class of
  equational theories.
  
\item We use a more general notion of constructor which allows
  pattern-matching on defined symbols and equations among
  constructors.
  
\item We relax the usual conditions on rewrite rules for ensuring the
  subject reduction property. By this way, we can eliminate some
  non-linearities in left-hand sides of rules and ease the confluence
  proof.
\end{lst}

%%%%%%%%%%%%%%%%%%%%%%%%%%%%%%%%%%%%%%%%%%%%%%%%%%%%%%%%%%%%%%%%%%%%%%%%%%%%%%
%%%%%%%%%%%%%%%%%%%%%%%%%%%%%%%%%%%%%%%%%%%%%%%%%%%%%%%%%%%%%%%%%%%%%%%%%%%%%%
%%%%%%%%%%%%%%%%%%%%%%%%%%%%%%%%%%%%%%%%%%%%%%%%%%%%%%%%%%%%%%%%%%%%%%%%%%%%%%

% directions for future work

%%%%%%%%%%%%%%%%%%%%%%%%%%%%%%%%%%%%%%%%%%%%%%%%%%%%%%%%%%%%%%%%%%%%%%%%%%%%%%
%%%%%%%%%%%%%%%%%%%%%%%%%%%%%%%%%%%%%%%%%%%%%%%%%%%%%%%%%%%%%%%%%%%%%%%%%%%%%%
%%%%%%%%%%%%%%%%%%%%%%%%%%%%%%%%%%%%%%%%%%%%%%%%%%%%%%%%%%%%%%%%%%%%%%%%%%%%%%

\section{Directions for future work}
\label{sec-future}

\begin{lst}{\bu}
\item In our conditions, we assume that the predicate symbols defined
  by rewrite rules containing quantifiers (``non-primitive'' predicate
  symbols) are defined by pattern-matching on free symbols only
  (``simple'' systems). It would be nice to be able to relax this
  condition.
  
\item Another important assumption is that the reduction relation
  $\a=\ar\cup\ab$ must be confluent. We will try to find sufficient
  conditions on $\cR$ in order to get the confluence of $\ar\cup\ab$.
  In the simply-typed $\la$-calculus, if $\cR$ is a first-order
  rewrite system then the confluence of $\cR$ is a sufficient
  condition \cite{breazu88lics}. But few results are known in the case
  of a richer type system or of higher-order rewriting.
  
\item Finally, we expect to extend this work with rewriting modulo
  some useful equational theories like associativity and
  commutativity, and also by allowing $\eta$-reductions in the type
  conversion rule.
\end{lst}

\newpage

\noindent
{\bf Acknowledgments~:} I would like to thank Daria Walukiewicz,
Gilles Dowek, Jean-Pierre Jouannaud and Christine Paulin for useful
comments on previous versions of this work.

%%% Local Variables: 
%%% mode: latex
%%% TeX-master: "main"
%%% End: 

% bibliography
\small
%\bibliographystyle{plain}
%\bibliography{\shortnames}

\begin{thebibliography}{10}

\bibitem{barbanera97jfp}
F.~Barbanera, M.~Fern\'andez, and H.~Geuvers.
\newblock Modularity of strong normalization in the algebraic-$\lambda$-cube.
\newblock {\em Journal of Functional Programming}, 7(6):613--660, 1997.

\bibitem{barendregt92book}
H.~Barendregt.
\newblock Lambda calculi with types.
\newblock In S.~Abramski, D.~Gabbay, and T.~Maibaum, editors, {\em Handbook of
  logic in computer science}, volume~2. Oxford University Press, 1992.

\bibitem{blanqui00rta}
F.~Blanqui.
\newblock Termination and confluence of higher-order rewrite systems.
\newblock In {\em Proc. of RTA'00\em, LNCS 1833}.

\bibitem{blanqui01thesis}
F.~Blanqui.
\newblock {\em Th\'eorie des {T}ypes et {R}\'e\'ecriture ({T}ype {T}heory and
  {R}ewriting)}.
\newblock PhD thesis, Universit\'e Paris-Sud (France), 2001.
\newblock Available at {\tt http://www.lri.fr/\~{}blanqui}. An english version
  will be available soon.

\bibitem{blanqui99rta}
F.~Blanqui, J.-P. Jouannaud, and M.~Okada.
\newblock The {C}alculus of {A}lgebraic {C}onstructions.
\newblock In {\em Proc. of RTA'99\em, LNCS 1631}.

\bibitem{blanqui01tcs}
F.~Blanqui, J.-P. Jouannaud, and M.~Okada.
\newblock Inductive-data-type systems.
\newblock {\em Theoretical Computer Science}, 277, 2001.

\bibitem{breazu88lics}
V.~Breazu-Tannen.
\newblock Combining algebra and higher-order types.
\newblock In {\em Proc. of LICS'88\em, IEEE Computer Society}.

\bibitem{breazu91tcs}
V.~Breazu-Tannen and J.~Gallier.
\newblock Polymorphic rewriting conserves algebraic strong normalization.
\newblock {\em Theoretical Computer Science}, 83(1):3--28, 1991.

\bibitem{coquand90lf}
T.~Coquand and J.~Gallier.
\newblock A proof of strong normalization for the {T}heory of {C}onstructions
  using a {K}ripke-like interpretation, 1990.
\newblock Paper presented at the 1st Int. Work. on Logical Frameworks but not
  published in the proceedings. Available at {\tt
  ftp://ftp.cis.upenn.edu/pub/papers/} {\tt gallier/sntoc.dvi.Z}.

\bibitem{coquand88ic}
T.~Coquand and G.~Huet.
\newblock The {C}alculus of {C}onstructions.
\newblock {\em Information and Computation}, 76(2--3):95--120, 1988.

\bibitem{dershowitz90book}
N.~Dershowitz and J.-P. Jouannaud.
\newblock Rewrite systems.
\newblock In J.~van Leeuwen, editor, {\em Handbook of Theoretical Computer
  Science}, volume~B, chapter~6. North-Holland, 1990.

\bibitem{dowek98trtpm}
G.~Dowek, T.~Hardin, and C.~Kirchner.
\newblock Theorem proving modulo.
\newblock Technical Report 3400, INRIA Rocquencourt (France), 1998.

\bibitem{dowek98types}
G.~Dowek and B.~Werner.
\newblock Proof normalization modulo.
\newblock In {\em Proc. of TYPES'98\em, LNCS 1657}.

\bibitem{dowek00note}
G.~Dowek and B.~Werner.
\newblock An inconsistent theory modulo defined by a confluent and terminating
  rewrite system, 2000.
\newblock Available at {\tt http://pauillac.inria.fr/\~{}dowek/}.

\bibitem{elan}
C.~Kirchner {\em et al\,}.
\newblock {ELAN}, 2000.
\newblock Available at {\tt http://elan.loria.fr/}.

\bibitem{cime}
C.~March\'e {\em et al\,}.
\newblock {CiME}, 2000.
\newblock Available at {\tt http://www.lri.fr/\~{}demons/cime.html}.

\bibitem{coq}
C.~Paulin {\em et al\,}.
\newblock {\em The {C}oq Proof Assistant Reference Manual Version 6.3.1}.
\newblock INRIA Rocquencourt (France), 2000.
\newblock Available at {\tt http://coq.inria.fr/}.

\bibitem{debruijn94book}
H.~Geuvers, R.~Nederpelt, and R.~de~Vrijer, editors.
\newblock {\em Selected Papers on Automath}, volume 133 of {\em Studies in
  Logic and the Foundations of Mathematics}.
\newblock North-Holland, 1994.

\bibitem{girard88book}
J.-Y. Girard, Y.~Lafont, and P.~Taylor.
\newblock {\em Proofs and Types}.
\newblock Cambridge University Press, 1988.

\bibitem{hsiang85ai}
J.~Hsiang.
\newblock Refutational theorem proving using term-rewriting systems.
\newblock {\em Artificial Intelligence}, 25:255--300, 1985.

\bibitem{jouannaud97tcs}
J.-P. Jouannaud and M.~Okada.
\newblock {A}bstract {D}ata {T}ype {S}ystems.
\newblock {\em Theoretical Computer Science}, 173(2):349--391, 1997.

\bibitem{jouannaud99lics}
J.-P. Jouannaud and A.~Rubio.
\newblock The {H}igher-{O}rder {R}ecursive {P}ath {O}rdering.
\newblock In {\em Proc. of LICS'99\em, IEEE Computer Society}.

\bibitem{klop93tcs}
J.~W. Klop, V.~van Oostrom, and F.~van Raamsdonk.
\newblock Combinatory reduction systems : introduction and survey.
\newblock {\em Theoretical Computer Science}, 121:279--308, 1993.

\bibitem{martinlof84book}
P.~Martin-L\"of.
\newblock {\em Intuitionistic type theory}.
\newblock Bibliopolis, Napoli, Italy, 1984.

\bibitem{mendler87thesis}
N.~P. Mendler.
\newblock {\em Inductive Definition in Type Theory}.
\newblock PhD thesis, Cornell University, United States, 1987.

\bibitem{paulin93tlca}
C.~Paulin-Mohring.
\newblock Inductive definitions in the system {C}oq - rules and properties.
\newblock In {\em Proc. of TLCA'93\em, LNCS 664}.

\bibitem{stefanova98thesis}
M.~Stefanova.
\newblock {\em Properties of Typing Systems}.
\newblock PhD thesis, Nijmegen University (Netherlands), 1998.

\bibitem{vandepol96thesis}
J.~van~de Pol.
\newblock {\em Termination of higher-order rewrite systems}.
\newblock PhD thesis, University of Utrecht, Nederlands, 1994.

\bibitem{oostrom94thesis}
V.~van Oostrom.
\newblock {\em Confluence for Abstract and Higher-Order Rewriting}.
\newblock PhD thesis, Vrije Universiteit, Netherlands, 1994.

\bibitem{walukiewicz00lfm}
D.~Walukiewicz.
\newblock Termination of rewriting in the {C}alculus of {C}onstructions.
\newblock In {\em Proc. of LFM'00}.

\bibitem{werner94thesis}
B.~Werner.
\newblock {\em Une Th\'eorie des {C}onstructions {I}nductives}.
\newblock PhD thesis, Universit\'e Paris VII, France, 1994.

\end{thebibliography}

% end document
\end{document}